%% file: main.tex
\documentclass[final,3p,times,twocolumn]{elsarticle}

\input{setup/preamble.tex}

\begin{document}

\input{sections/01_title.tex}
\input{sections/02_abstract.tex}

\maketitle

\input{sections/03_introduction.tex}
\input{sections/04_forex_overview.tex}
\input{sections/05_data.tex}
\input{sections/06_time_scale.tex}
\input{sections/07_response_functions.tex}
\input{sections/08_spread_impact.tex}

\input{sections/09_conclusion.tex}

\input{sections/10_paper_contributions.tex}

\appendix
\input{sections/11_appendix_A.tex}

\bibliographystyle{unsrt}
\bibliography{bib}

\end{document}

%% file: setup/preamble.tex
\usepackage{latexsym}
\usepackage{graphicx}
\usepackage{hyperref} 
\usepackage[english]{babel}
\usepackage[utf8]{inputenc}
\usepackage{xcolor}
\usepackage{amsmath}
\usepackage{amstext}
\usepackage{url}
\usepackage[english]{babel}
\usepackage{booktabs}
\usepackage{threeparttable}
\usepackage[title]{appendix}
\usepackage{array}

\newcolumntype{P}[1]{>{\centering\arraybackslash}p{#1}}

\hypersetup{%
   	pdfpagelabels=true,%
	plainpages=false,%
	pdfauthor={J. Henao-Londono et al.},%
	pdftitle={Response functions and spread impact in foreign exchange markets},%
	pdfsubject={Complex Systems},%
	bookmarksnumbered=true,%
	colorlinks=true,%
	citecolor=blue,%
	filecolor=black,%
	linkcolor=blue,
	urlcolor=blue,%
	pdfstartview=FitH%
}

%% file: sections/01_title.tex
\title{Foreign exchange markets: price response and spread impact}
\author[ude]{Juan C. Henao-Londono \fnref{henao}}
\ead{juan.henao-londono@uni-due.de}

\author[ude]{Thomas Guhr \fnref{guhr}}
\ead{thomas.guhr@uni-due.de}

\affiliation[ude]{
        organization={Fakultät für Physik, Universität Duisburg-Essen},
        addressline={Lotharstraße 1},
        city={Duisburg},
        postcode={47048},
        state={NRW},
        country={Germany}}

\cortext[cor1]{Corresponding author}

%% file: sections/02_abstract.tex
\begin{abstract}
      We carry out a detailed large-scale data analysis of price response
      functions in the spot foreign exchange market for different years and
      different time scales. Such response functions provide quantitative
      information on the deviation from Markovian behavior. The price response
      functions show an increase to a maximum followed by a slow decrease as
      the time lag grows, in trade time scale and in physical time scale, for
      all analyzed years. Furthermore, we use a price increment point (pip)
      bid-ask spread definition to group different foreign exchange pairs and
      analyze the impact of the spread in the price response functions. We find
      that large pip spreads have a stronger impact on the response. This is
      similar to what has been found in stock markets.
\end{abstract} 

\begin{keyword}
      Econophysics \sep Complex systems \sep Statistical physics \sep
      Price response function \sep Spread impact \sep Foreign exchange market
\end{keyword}

%% file: sections/03_introduction.tex
\section{Introduction}\label{sec:introduction}

A major objective of data driven research on complex systems is the
identification of generic or universal statistical behavior. The tremendous
success of thermodynamics and statistical mechanics serves as an inspiration
when continuing this quest in complex systems beyond traditional physics.
Particularly interesting are large complex systems which consist of similar,
yet clearly distinguishable complex subsystems. Financial markets, for example,
have well defined subsystems as foreign exchange markets, stock markets, bond
markets, among others. The degree of universality found in one particular
subsystem can then be assessed if this type of universality is also seen in
another subsystem. If applicable, useful information on the impact of specific
system features on this universality may then be inferred.

In spite of the considerable interest, a thorough statistical analysis of
the microstructure in foreign exchange markets was hampered by limited access
to data. This changed, and nowadays such data analyses are possible down to the
level of ticks and over long time scales.

Here, we carry out such a study for finance, because a tremendous amount of
data is available \cite{physicists_contribution}. Markets may be viewed as
macroscopic complex systems with an internal microscopic structure that is to a
large extent accessible by big data analysis \cite{complex_markets}.
Stock markets and foreign exchange markets are clearly distinct, but share many
common features. In previous analyses, we studied response functions in stock
markets to shed light on non-Markovian behavior. Here, we extend that to the
spot foreign exchange markets. To our surprise, we did not find such an
investigation in the literature. Hence, we believe that this study is a
rewarding effort. It helps to examine the behavior of the functions applied to
the foreign exchange market and it is suitable to compare the similarities and
differences to other markets.

The foreign exchange market has attracted a lot of attention in the last 20
years. Electronic trading has changed an opaque market to a fairly transparent
one with transaction costs that are a fraction of their former level. The large
amount of data that is now available to the public makes possible different
kinds of data analysis. Intense research is currently carried out in different
directions
\cite{forex_liquidity,info_forex,intraday_forex,forex_structure,teach_spread,forex_microstructure,electronic_forex,forex_algorithmic,curr_speculation,patterns_forex,eur_change_forex,spread_competition,political_forex,forex_volatility,local_forex,forex_inefficiency}.

McGroarty et al. \cite{micro_eff} found that smaller volumes cause larger
bid-ask spreads for technical reasons related to the measurement, whereas Hau
et al. \cite{eur_change_forex,eur_int_curr} claim that larger bid-ask spreads
caused smaller volumes due to the traders' behavior.

Burnside et al. \cite{curr_speculation} found the spreads to be between two and
four times larger for emerging market currencies than for developed country
currencies. According to Huang and Masulis \cite{spread_competition}, bid-ask
spreads increase when the foreign exchange market volatility increases, and
decrease when the competition between the dealers increases. Ding and Hiltrop
\cite{electronic_forex} showed that the Electronic Broking Services (EBS)
reduces spreads significantly, but dealers with information advantage tend to
quote relatively wider spreads. King \cite{spread_futures} analyzed the foreign
exchange futures market and observed that the number of transactions is
negatively related with bid-ask spread, whereas volatility in general is
positively related. Serbinenko and Rachev \cite{intraday_forex} focus on the
three major market characteristics, namely efficiency, liquidity and
volatility, and found that the market is efficient in a weak form. Menkhoff and
Schmeling \cite{local_forex} used orders from the Russian interbank for Russian
rouble/US dollar rate. They analyzed the price impact in different regions of
Russia, and found that regions that are centers of political and financial
decision making have high permanent price impact.

Price response functions are a powerful tool to obtain dynamical information
because they measure price changes implied by execution of market orders.
Specifically, they measure how a buy or sell order at time $t$ influences on
average the price at a later time $t + \tau$. It was shown in different works
\cite{components_spread_tokyo,dissecting_cross,r_walks_liquidity,subtle_nature,Bouchaud_2004,theory_market_impact,my_paper_response_financial,Wang_2016_avg,Wang_2016_cross}
that the price response functions increase to a maximum and then slowly
decrease as the time lag grows.

Little is known about price response functions or related quantities in the
foreign exchange markets
\cite{forex_liquidity,forex_volatility,response_funct_fx}. Melvin and Melvin
\cite{forex_volatility} simulate their proposed model for different foreign
exchange markets region to analyze the impact of a one-standard-deviation shock
using impulse response functions. The general pattern of response was a fairly
steep drop over the first couple of days followed by a few days of gradual
decline until the response is not statistically different from zero. Mancini et
al. \cite{forex_liquidity} model the price impact and return reversal to
analyze liquidity. Their model predicts that more liquid assets should exhibit
narrower spreads and lower price impact.

To the best of our knowledge, no large-scale data analysis of response
functions for the spot foreign exchange market has been carried out. Response
functions are important observables as they give information on non-Markovian
behavior. It is the purpose of the present study to close this gap. Based on a
series of detailed empirical results obtained on trade by trade data, we show
that the price response functions in the foreign exchange markets behave
qualitatively similar as the ones in correlated stocks markets. We consider
different time scales, years and currency pairs to compute the price response
functions. Finally, we shed light on the spread impact in the response
functions for foreign exchange pairs. We use a pip bid-ask spread definition to
group different foreign exchange pairs and show that large pip spreads have a
stronger impact on the response. To facilitate the reproduction of our results,
the source code for the data analysis is available in Ref. \cite{code}.

The paper is organized as follows: in Sect. \ref{sec:forex_overview} we
introduce the foreign exchange market. In Sect. \ref{sec:data_set} we present
our data set of spot foreign exchange pairs and briefly describe the physical
and trade time scale. We define the time scale to be used in Sect.
\ref{sec:time_scale}, and compute the price response functions for foreign
exchange pairs in Sect.  \ref{sec:response_functions}. In Sect.
\ref{sec:spread_impact} we show how the spread impact the values of the
response functions. Our conclusions follow in Sect. \ref{sec:conclusion}.

%% file: sections/04_forex_overview.tex
\section{Foreign exchange market overview}\label{sec:forex_overview}

In Sect. \ref{subsec:forex_market} we describe the basic characteristics of the
foreign exchange market. In Sect. \ref{subsec:key_concepts} we establish the
fundamental quantities used in the price response definitions.

\subsection{Foreign exchange market}\label{subsec:forex_market}

The foreign exchange market is a 24-hour global decentralized or
over-the-counter (OTC) market for the trading of currencies closing only on the
weekends.
The foreign exchange market is the most volatile, liquid and largest of all
financial markets
\cite{forex_liquidity,info_forex,intraday_forex,forex_structure,teach_spread,forex_market_micro,book_forex,book_forex_2,book_forex_3},
and it has a paramount importance for the world economy. It affects employment,
inflation, international capital flows, among others \cite{forex_structure}.
The major participants trading in this market include governments, central
banks, global funds, retail clients and corporations
\cite{book_forex_2,book_forex_3}. Trading of currency in the foreign exchange
market involves the purchase and sale of two currencies at the same time
\cite{book_forex,book_forex_2,book_forex_3}. The value of one of the currencies
in that pair is relative to the value of the other. The price one currency can
be exchanged with another currency is the foreign exchange rate. The foreign
exchange market is a closed system. As one value increases another value has
to decrease. All foreign exchange rates cannot appreciate, in contrast to the
stock market \cite{book_forex,book_forex_3}.

Depending on the country, the currencies can be ``free float'' or
``fixed float''. Free-floating currencies relative value is determined by
free-market forces. Some example of free-floating currencies include the U.S.
dollar, Japanese yen and Colombian peso. On the other hand, a fixed float is
where a government through the central bank set the currency's relative value
to other currencies, usually by pegging it to some standard. Examples of fixed
floating currencies include the Chinese Yuan and the Indian Rupee
\cite{book_forex}. In our case, we only use free float currencies.

In the foreign exchange market, the trading day begins in Australia and Asia.
Then the markets in Europe open and finally the markets in America
\cite{forex_structure,forex_market_micro,book_forex_2,book_forex_3}. As the
market close time in New York overlaps the market open time in Australia and
Asia, the markets do not formally close during the week. Thus, using the New
York time as reference, the market opens on Sunday at 19h00 and closes on
Friday at 17h00. London, New York and Tokyo are the largest trading centers of
foreign exchange trading \cite{book_forex_4}

Currency markets are divided into spot market, forward, future, currency swaps
and currency options \cite{book_forex_2,book_forex_3,book_forex_4}. In our work
we particularly focus on the spot market, where as his name suggest, the trades
are settled on the spot \cite{book_forex,book_forex_3}. In a spot market, as
the ccurency transactions are carried in the OTC markets, information
concerning open interest and volume is unavailable. The transactions in this
market represent up to the 40\% of the total market transactions in the foreign
exchange market. The most traded currencies in the spot market are the U.S.
dollar, euro, Japanese yen, British pound and Swiss franc \cite{book_forex}.

In general, three categories of currency pairs are defined: majors, crosses,
and exotics. The ``major'' foreign exchange currency pairs are the most
frequently traded currencies that are paired with the U.S. dollar. The
``crosses'' are those majors pairs paired between them and that exclude the
U.S. dollar. Finally, the ``exotic'' pairs usually consist of a major currency
alongside a thinly traded currency or an emerging market economy currency. The
majors are the most liquid pairs, in contrast with the exotics, who can be much
more volatile. In this work, we will refer as the ``major currency pairs'' to
the pairs of most traded currencies paired with the U.S. dollar, including the
so called commodity currencies: Canadian dollar, Australian dollar and New
Zealand dollar. The pairs and their corresponding symbol can be seen in Table
\ref{tab:majors}.

\begin{table}[htbp]
\centering
\begin{threeparttable}
\caption{Analyzed currency pairs.}
\begin{tabular*}{\columnwidth}{P{5cm}P{3cm}}
\toprule
\bf{Currency pair} & \bf{Symbol} \tabularnewline
\midrule
euro/U.S dollar& EUR/USD \tabularnewline
British pound/U.S. dollar& GBP/USD \tabularnewline
Japanese yen/U.S. dollar& JPY/USD \tabularnewline
Australian dollar/U.S. dollar& AUD/USD \tabularnewline
U.S. dollar/Swiss franc& USD/CHF \tabularnewline
U.S. dollar/Canadian dollar& USD/CAD \tabularnewline
New Zealand dollar/U.S. dollar& NZD/USD \tabularnewline
\bottomrule
\end{tabular*}
\label{tab:majors}
\end{threeparttable}
\end{table}

The term pip (Price Increment Point) is commonly used in the foreign exchange
market instead of tick. The precise definition of a pip is a matter of
convention. Usually, it refers to the incremental value in the fifth non-zero
digit position from the left. It is not related to the position of the decimal
point. For example, one pip in the exchange rate USD/JPY of 124.21 would be
0.01, while one pip for EUR/USD of 1.1021 would be 0.0001
\cite{forex_structure,micro_eff,forex_market_micro,book_forex_3,order_flow_forex}.

Compared with other markets like the stock market, there are some key
characteristics that differentiate the foreign exchange market. There are fewer
rules, there are no clearing houses and central bodies that oversee the market.
The investors will not have to pay fees or commissions as on another markets.
It is possible to trade at any time of day and regarding the risk and reward,
it is possible to get in and out whenever the investor want. In the foreign
exchange market, the bid-ask spread is the only transaction cost
\cite{book_forex_2}.

\subsection{Key concepts}\label{subsec:key_concepts}

In spot foreign exchange markets, orders are executed at the best available buy
or sell price. Orders often fail to result in an immediate transaction, and are
stored in a queue called the limit order book
\cite{forex_structure,forex_market_micro,stat_prop,predictive_pow,intro_market_micro,prop_order_book}.
The order book is visible for all traders and its main purpose is to ensure
that all traders have the same information on what is offered on the market.
For a detailed description of the operation of the markets, we suggest to see
Ref. \cite{my_paper_response_financial}.

At any given time there is a best (lowest) offer to sell with price
$a\left(t\right)$, and a best (highest) bid to buy with price $b\left(t\right)$
\cite{subtle_nature,book_forex,prop_order_book,account_spread,limit_ord_spread,stat_theory}.
The price gap between them is called the spread
$s\left(t\right) = a\left(t\right)-b\left(t\right)$
\cite{teach_spread,subtle_nature,Bouchaud_2004,book_forex,account_spread,stat_theory,large_prices_changes,market_digest,em_stylized_facts}.
Spreads are significantly positively related to price. Currencies with more
liquidity tend to have lower spreads
\cite{components_spread_tokyo,account_spread,effects_spread,components_spread}.
In spot foreign exchange markets, the existing spread in any currency will vary
depending on the currency trader, the currency being traded and the conditions
in the market. Although the foreign exchange market is often cited as the
world's largest financial market, this description fails to consider the
considerable differences in trading volume and liquidity across different
currency pairs \cite{forex_microstructure,book_forex_2}. These differences can
be directly seen in the spread. The spread will tend to increase for currencies
that do not generate a large volume of trading \cite{book_forex_2}.
Furthermore, the bid-ask spread is directly related with the transaction costs
to the dealer \cite{teach_spread,spread_futures,book_forex_2}.

As we have the quotes prices in the data, we need to infer the trade price. We
consider a basic definition of the price given by
\cite{forex_liquidity,patterns_forex,political_forex}. The average of the best
ask and the best bid is the midpoint price, which is defined as
\cite{teach_spread,subtle_nature,Bouchaud_2004,my_paper_response_financial,prop_order_book,stat_theory,large_prices_changes,em_stylized_facts}
\begin{equation}
    m \left(t\right) = \frac{a\left(t\right) + b\left(t\right)}{2}.
\end{equation}
Price changes are typically characterized as returns. Using the midpoint price
$m\left( t\right)$ of a currency pair at time $t$, the return
$r\left(t, \tau\right)$, at time $t$ and time lag $\tau$ is simply the relative
variation of the price from $t$ to $t + \tau$
\cite{subtle_nature,empirical_facts,asynchrony_effects_corr,tick_size_impact,causes_epps_effect,non_stationarity},
\begin{equation}\label{eq:midpoint_price_return}
    r\left(t,\tau\right) = \frac{m\left(t+\tau\right)-m\left(t\right)}
    {m\left(t\right)}.
\end{equation}
The distribution of returns is strongly non-Gaussian and its shape continuously
depends on the return period $\tau$. Small $\tau$ values have fat tails return
distributions \cite{subtle_nature}. The trade signs are defined for general
cases as
\begin{equation}\label{eq:trade_sign_general}
    \varepsilon\left(t\right)=\text{sign}\left(S\left(t\right)
    -m\left(t-\delta\right)\right),
\end{equation}
where $\delta$ is a positive time increment. Hence we have
\begin{equation}\label{eq:trade_sign_results}
    \varepsilon\left(t\right)=\left\{
    \begin{array}{cc}
    +1, & \text{If } S\left(t\right)
    \text{ is higher than the last } m\left( t \right)\\
    -1, & \text{If } S\left(t\right)
    \text{ is lower than the last } m\left( t \right)
    \end{array}\right. .
\end{equation}
Here, $\varepsilon(t) = +1$ indicates that the trade was triggered by a market
order to buy and a trade triggered by a market order to sell yields
$\varepsilon(t) = -1$
\cite{subtle_nature,Bouchaud_2004,spread_changes_affect,quant_stock_price_response,order_flow_persistent}.
In our implementation of the price response functions we need a more specific
definition of trade signs. We give a deeper explanation of trade signs
depending on the time scale in Sect. \ref{sec:time_scale}.

The main objective of this work is to analyze the price response functions. In
general we define the price response functions in a foreign exchange market as
\begin{equation}\label{eq:response_general}
    R^{\left(\textrm{scale}\right)}_{i}\left(\tau\right)=\left\langle
    r^{\left(\textrm{scale}\right)}_{i}\left(t-1, \tau\right)
    \varepsilon^{\left(\textrm{scale}\right)}_{i} \left(t\right)\right\rangle
    _{\textrm{average}},
\end{equation}
where the index $i$ corresponds to currency pairs in the market,
$r^{\left(\textrm{scale}\right)}_{i}$ is the return of the pair $i$ in a time
lag $\tau$ in the corresponding scale and
$\varepsilon^{\left(\textrm{scale}\right)}_{i}$ is the trade sign of the pair
$i$ in the corresponding scale. The superscript scale refers to the time scale
used, whether physical time scale ($\textrm{scale} = \textrm{p}$) or trade time
scale ($\textrm{scale} = \textrm{t}$). Finally, The subscript average refers to
the way to average the price response, whether relative to the physical time
scale ($\textrm{average} = P$) or relative to the trade time scale
($\textrm{average} = T$).

For correlated financial markets, the price response function increases to a
maximum and then slowly decreases. This result is observed empirically in trade
time scale and in physical time scale
\cite{my_paper_response_financial,Wang_2016_avg}.

%% file: sections/05_data.tex
\section{Data set}\label{sec:data_set}

The spot foreign exchange financial data was obtained from
\href{www.histdata.com}{HistData.com}. We use a tick-by-tick database in
generic ASCII format for different years and currency pairs. This tick-by-tick
data is sampled for each transaction. The data comprises the date time stamp
(YYYYMMDD HHMMSSNNN), the best bid and best ask quotes prices in the Eastern
Standard Time (EST) time zone. With both best bid and best ask quotes it is
easy to compute the pip bid-ask spread of the data. No information about the
size of each transaction is provided. Also, the identity of the participants is
not given. Furthermore, trading volumes in spot foreign exchange market are not
aggregated and the only volumes that are possible to find are the Broker
Specific Volumes. Therefore, the data provider decided to remove the volume
information from the delivered data.

Our aims is to compute the price response functions in two different
time scales: trade time scale and physical time scale. We give a complete
description on how the scales are defined in Sect.  \ref{sec:time_scale}.
Regarding the data for this definitions, for the trade time scale we use the
data as it is, considering that it is sampled for each transaction. On the
physical time scale, for each exchange rate, we process the irregularly spaced
raw data to construct second-by-second price series, each containing 86,400
observations per day. For every second, the midpoint of best bid and ask quotes
are used to construct one-second log-returns.

Another goal in this paper is to compare the price response functions in
different calendar years to see the differences and similarities along time.
To analyze the price response functions in Sect. \ref{sec:response_functions},
we select the seven major currency pairs (see Table \ref{tab:majors}) in three
different years: 2008, 2014 and 2019.

Additionally, we analyze the pip bid-ask spread impact in price response
functions (Sect. \ref{sec:spread_impact}). We select 46 currency pairs in three
different years (2011, 2015 and 2019). The selected pairs are listed in
\ref{app:fx_pairs_spread}.

The selection of the calendar years to be analyzed was made considering the
availability of the data, the completeness of the time series and the option to
have a constant gap between the years.

In order to avoid overnight effects and any artifact due to the opening and
closing of the foreign exchange market, we systematically discard the first
ten and the last ten minutes of trading in a given week
\cite{Bouchaud_2004,my_paper_response_financial,Wang_2016_cross,large_prices_changes,spread_changes_affect}.
Therefore, we only consider trades of the same week from Sunday 19:10:00 to
Friday 16:50:00 New York local time. We will refer to this interval of time as
the ``market time".

%% file: sections/06_time_scale.tex
\section{Time scale}\label{sec:time_scale}

In Sect. \ref{subsec:time_definition} we describe the physical time scale and
the trade time scale. In Sect. \ref{subsec:trade_time} and Sect.
\ref{subsec:physical_time} we define the trade and the physical time scales,
respectively.

\subsection{Time definition}\label{subsec:time_definition}

Due to the nature of the data, they are several options to define time for
analyzing data. In general, the time series are labeled in calendar time
(hours, minutes, seconds, milliseconds). Moreover, tick-by-tick data available
on financial markets all over the world is time stamped up to the millisecond,
but the order of magnitude of the guaranteed precision is much larger, usually
one second or a few hundreds of milliseconds
\cite{market_digest,empirical_facts}. In several papers are used different time
definitions (calendar time, physical time, event time, trade time, tick time)
\cite{empirical_facts,sampling_returns,market_making}. The spot foreign
exchange market data used in the analysis only has the quotes. In consequence,
we have to infer the trades during the market time. As we have tick-by-tick
resolution, we can use either trade time scale or physical time scale.

The trade time scale increases by one unit each time a transaction happens,
which in our case is every time the quotes change. The advantage of this count
is that limit orders far away in the order book do not increase the time by one
unit. The main outcome of trade time scale is its ``smoothing" of data and the
aggregational normality \cite{empirical_facts}.

The physical time scale is increased by one unit each time a second passes.
This means that computing the responses in this scale involves sampling
\cite{Wang_2016_cross,sampling_returns}, which has to be done carefully when
dealing for example with several stocks with different liquidity. This sampling
is made in the trade signs and in the midpoint prices.

We use these two definitions of time scale to compute the price response
function. Between these two scales, there is not a judgment which is better or
worse. Their use directly depend on the application. Thus, our aim is to
present how the price response function behaves under these two different time
scales.

\subsection{Trade time scale}\label{subsec:trade_time}

As a first approximation, we use the trade sign classification in trade time
scale proposed in Ref. \cite{Wang_2016_cross} and used in Refs.
\cite{my_paper_response_financial,Wang_2016_avg,Wang_2017,Wang_2018_copulas}
that reads
\begin{equation}\label{eq:trade_signs_trade}
    \varepsilon^{\left(\textrm{t}\right)}\left(t,n\right)=\left\{
    \begin{array}{cc}
    \text{sgn}\left(m\left(t,n\right)-m\left(t,n-1\right)\right),
    & \text{if }\\ m\left(t,n\right) \ne m\left(t,n-1\right)\\
    \varepsilon^{\left(\textrm{t}\right)}\left(t,n-1\right),
    & \text{otherwise}
    \end{array}\right..
\end{equation}
Here, $\varepsilon^{\left(\textrm{t}\right)}\left( t,n \right) = +1$ implies a
trade triggered by a market order to buy, and a value
$\varepsilon^{\left(\textrm{t}\right)}\left( t,n \right) = -1$ indicates a trade
triggered by a market order to sell.

In the second case of Eq. (\ref{eq:trade_signs_trade}), if two consecutive
trades with the same trading direction do not exhaust all the available volume
at the best quote, the trades would have the same price, and they will thus
have the same trade sign.

With this classification we obtain trade signs for every single trade in the
data set. According to Ref. \cite{Wang_2016_cross}, the average accuracy of the
classification is $85\%$ for the trade time scale.

\subsection{Physical time scale}\label{subsec:physical_time}

We use the trade sign definition in physical time scale proposed in Ref.
\cite{Wang_2016_cross} and used in Refs.
\cite{Wang_2016_avg,Wang_2017}, that depends on the classification in
Eq. (\ref{eq:trade_signs_trade}) and reads
\begin{equation}\label{eq:trade_signs_physical}
    \varepsilon^{\left(\textrm{p}\right)}\left(t\right)=\left\{
    \begin{array}{cc}
    \text{sgn}\left(\sum_{n=1}^{N\left(t\right)}
    \varepsilon^{\left(\textrm{t}\right)} \left(t,n\right)\right),
    & \text{If }N \left(t\right)>0\\
    0, & \text{If }N\left(t\right)=0
    \end{array}\right. ,
\end{equation}
where $N \left(t \right)$ is the number of trades in an interval of one second.
Here, $\varepsilon^{\left(\textrm{p}\right)}\left( t \right) = +1$ implies that
the majority of trades in the second $t$ are triggered by a market order to buy,
and a value $\varepsilon^{\left(\textrm{p}\right)}\left( t \right) = -1$
indicates a majority of sell market orders. In this definition, there are two
ways to obtain $\varepsilon^{\left(\textrm{p}\right)}\left( t \right) = 0$.
First, there are no trades in a particular second and thus no trade sign.
Second, the sum of the trade signs in a given second amounts to zero,
indicating an exact balance of buy and sell market orders.

Market orders show opposite trade directions as compared to limit orders
executed simultaneously. An executed sell limit order corresponds to a
buyer-initiated market order. An executed buy limit order corresponds to a
seller-initiated market order. In this case we do not compare every single
trade sign in a second, but the net trade sign obtained for every second with
the definition, see Eq. (\ref{eq:trade_signs_physical}). According to Ref.
\cite{Wang_2016_cross}, this definition has an average accuracy up to $82\%$ in
the physical time scale.

%% file: sections/07_response_functions.tex
\section{Price response functions}\label{sec:response_functions}

In Sect. \ref{subsec:response_function_trade} we analyze the responses
functions in trade time scale and in Sect.
\ref{subsec:response_function_physical} we analyze the responses functions in
physical time scale.

\subsection{Response functions on trade time scale}
\label{subsec:response_function_trade}

The price response function in trade time scale is defined as
\cite{my_paper_response_financial}
\begin{equation}\label{eq:response_functions_trade_scale_general}
    R^{\left(\textrm{t}\right)}_{i}\left(\tau\right)=\left\langle
    r^{\left(\textrm{t}\right)}_{i}\left(t-1,\tau \right)
    \varepsilon_{i}^{\left(\textrm{t}\right)}
    \left(t, n\right)\right\rangle _{T}.
\end{equation}
To compute the response functions on trade time scale, we use both, the trade
signs and the returns from the tick-by-tick original data during a week in
market time. Then, the response is averaged by the number of trades.

\begin{figure}[htbp]
    \centering
    \includegraphics[width=\columnwidth]
    {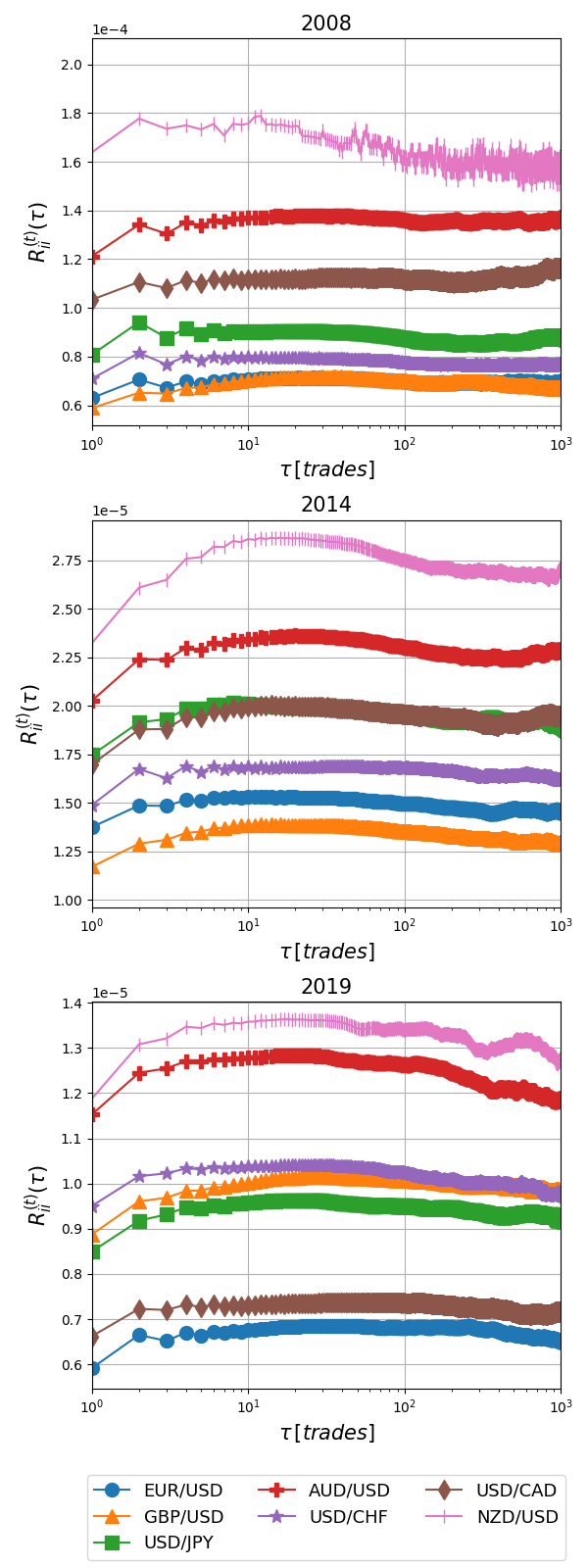}
    \caption{Price response functions
             $R^{\left(\textrm{t}\right)}_{i}\left(\tau\right)$ versus time
             lag $\tau$ on a logarithmic scale in trade time scale for the
             years 2008 (top), 2014 (middle) and 2019 (bottom).}
    \label{fig:response_function_trade_scale}
\end{figure}

The results of Fig. \ref{fig:response_function_trade_scale} show the
price response functions of the seven foreign exchange major pairs used in the
analysis (see Table \ref{tab:majors}) for three different years. The results
found for all the years are entirely in line with price responses seen in other
financial markets, particularly with correlated financial markets. The response
functions have an initial increasing trend to a maximum, that flattens out and
saturates at some level, and eventually slowly decrease. This shape is
explained by an initial increase caused by autocorrelated transaction flow. The
flattening out is due to the market liquidity adapting to this flow and assuring
diffusive prices \cite{EMH_lillo}. For our selected pairs, a time lag of
$\tau = 10^{3} $ trades is enough to see an increase to a maximum followed by a
decrease. Thus, the trend in the price response functions is eventually
reversed. The response signal is much more noisier in the year 2008 for the
first seconds in the time lag. This behavior is because of the smaller amount
of data of the corresponding year. In general, more data was recorded in recent
years than in past years. In the three years analyzed, the more liquid currency
pairs have a smaller response in comparison with the non-liquid pairs.  The
strength of the response function varies from one year to the other. In 2008
the strength of the signal was one order of magnitude stronger than the
response in 2014, but the signals in 2014 have approximately twice the strength
of the signals of 2019. This behavior can be explained by the fact that in
recent times algorithm trading has been used intensively. Thus, many more
trades were carried out in the last years, which means, the impact of each
trade is reduced, and then the response functions tend to decrease compared
with previous years.

\subsection{Response functions on physical time scale}
\label{subsec:response_function_physical}

One important detail to compute the price response function on physical time
scale is to define how the averaging of the function will be made, because the
response functions highly differ when we include or exclude
$\varepsilon^{\left(\textrm{p}\right)}_j \left( t\right) = 0$
\cite{Wang_2016_cross}. The price responses including
$\varepsilon^{\left(\textrm{p}\right)}_j \left( t\right) = 0$ are weaker than
the excluding ones due to the omission of direct influence of the lack of
trades. However, either including or excluding
$\varepsilon^{\left(\textrm{p}\right)}_j \left( t\right) = 0$ does not change
the trend of price reversion versus the time lag, but it does affect the
response function strength \cite{Wang_2016_avg}. For a deeper analysis of the
influence of the term
$\varepsilon^{\left(\textrm{p}\right)}_j \left( t\right) = 0$ in price response
functions, we suggest reviewing Refs. \cite{Wang_2016_avg,Wang_2016_cross}. We
will only take into account the price response functions excluding
$\varepsilon^{\textrm{p}}_j \left( t\right) = 0$.

We define the price response functions on physical time scale, using
the trade signs and the returns sampled in seconds from the original data on
physical time scale. The price response function on physical time scale is
defined as \cite{my_paper_response_financial}
\begin{equation}\label{eq:response_functions_time_scale_general}
    R^{\left(\textrm{p}\right)}_{i}\left(\tau\right)=\left\langle
    r^{\left(\textrm{p}\right)}_{i}\left(t-1, \tau\right)
    \varepsilon_{i}^{\left(\textrm{p}\right)} \left(t\right)\right\rangle _{P}
\end{equation}
\begin{figure}[htbp]
    \centering
    \includegraphics[width=\columnwidth]
    {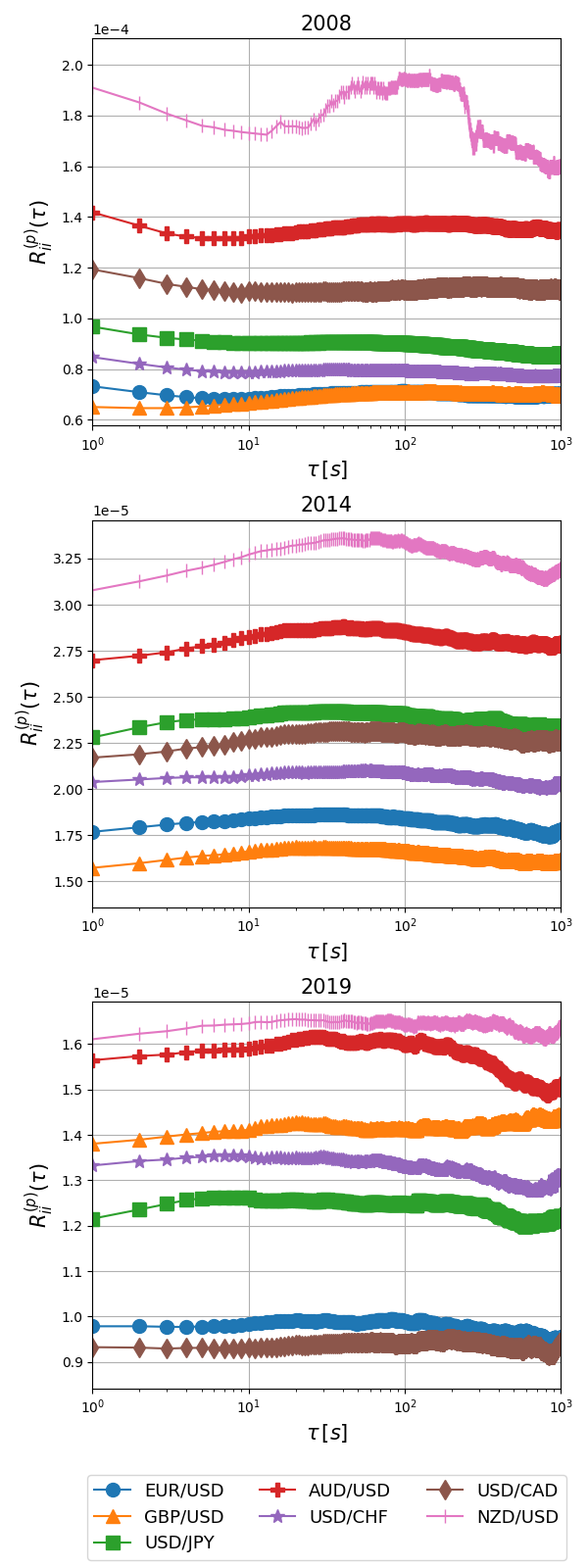}
    \caption{Price response functions
             $R^{\left(\textrm{p}\right)}_{i}\left(\tau\right)$ excluding
             $\varepsilon^{\left(\textrm{p}\right)}_{i}\left(t\right) = 0$ versus time
             lag $\tau$ on a logarithmic scale in physical time scale for the
             years 2008 (top), 2014 (middle) and 2019 (bottom).}
    \label{fig:response_function_physical_scale}
\end{figure}
The results shown in Fig. \ref{fig:response_function_physical_scale} are the
price response functions on physical time scale for three different years. The
results show approximately the same behavior observed in currency exchange
pairs in trade time scale, and in correlated financial markets, where we can
see that an increase to a maximum is followed by a decrease. Thus again, the
trend in the price responses is eventually reversed. An exception occurs in the
year 2008, where the response at short time lags seems to decrease, to then
start to slightly increase, and finally it decreases again.

The price response functions on physical time scale are smoother than the
responses on trade time scale. As we reduce from trade data all the returns and
trade signs in one second to one data point on physical time scale, and as this
sampling gives the same weight to every data point, the curves look smoother.

Compared with the response functions on trade time scale, the strength of the
signal of the response functions on physical time scale are similar in
magnitude in the corresponding years. Thus, the strength of the signal in 2008
for trade time scale is similar to the strength of the signal in 2008 for
physical time scale, and so on. This behavior is different to the one presented
in correlated financial markets, where the results differ about a factor of
two depending on the time scale.

On physical time scale, we can see that the liquid pairs have a smaller price
response compared with non-liquid pairs. Therefore, the price response of a
foreign exchange pair with large activity is smaller to the small impact of
each trade. Also, the older the response, the stronger the signal. We consider
the same argument of algorithm trading to explain why the signals in recent
years are weaker than in older years.

%% file: sections/08_spread_impact.tex
\section{Spread impact in price response functions}\label{sec:spread_impact}

To analyze the spread impact in price response functions, we use 47 foreign
exchange pairs from three different years (\ref{app:fx_pairs_spread}).
As we showed in Sect. \ref{sec:forex_overview}, due to the difference in the
position of the decimal points in the price between foreign exchange pairs, to
compare them we need to introduce a ``scaling factor'' with the purpose of
bringing the pip to the left of the decimal point. For example, the scaling
factor for the USD/JPY is $100$ and the one for the EUR/USD is $10000$.

The pip bid-ask spread is defined as \cite{micro_eff}
\begin{equation}
    s_{\textrm{pip}} = \left(a\left(t\right) - b\left(t\right)\right) \cdot
    \textrm{scaling factor}.
\end{equation}
With the $s_{\textrm{pip}}$ we can group the foreign exchange pairs and check
how the average strength of the price response functions on trade time scale
and physical time scale behave. For each pair we compute the pip bid-ask spread
in every trade along the market time. Then we average the spread during the
trade weeks in the different years. With this value we group the foreign
exchange pairs.

\begin{figure*}[htbp]
    \centering
    \includegraphics[width=\textwidth]{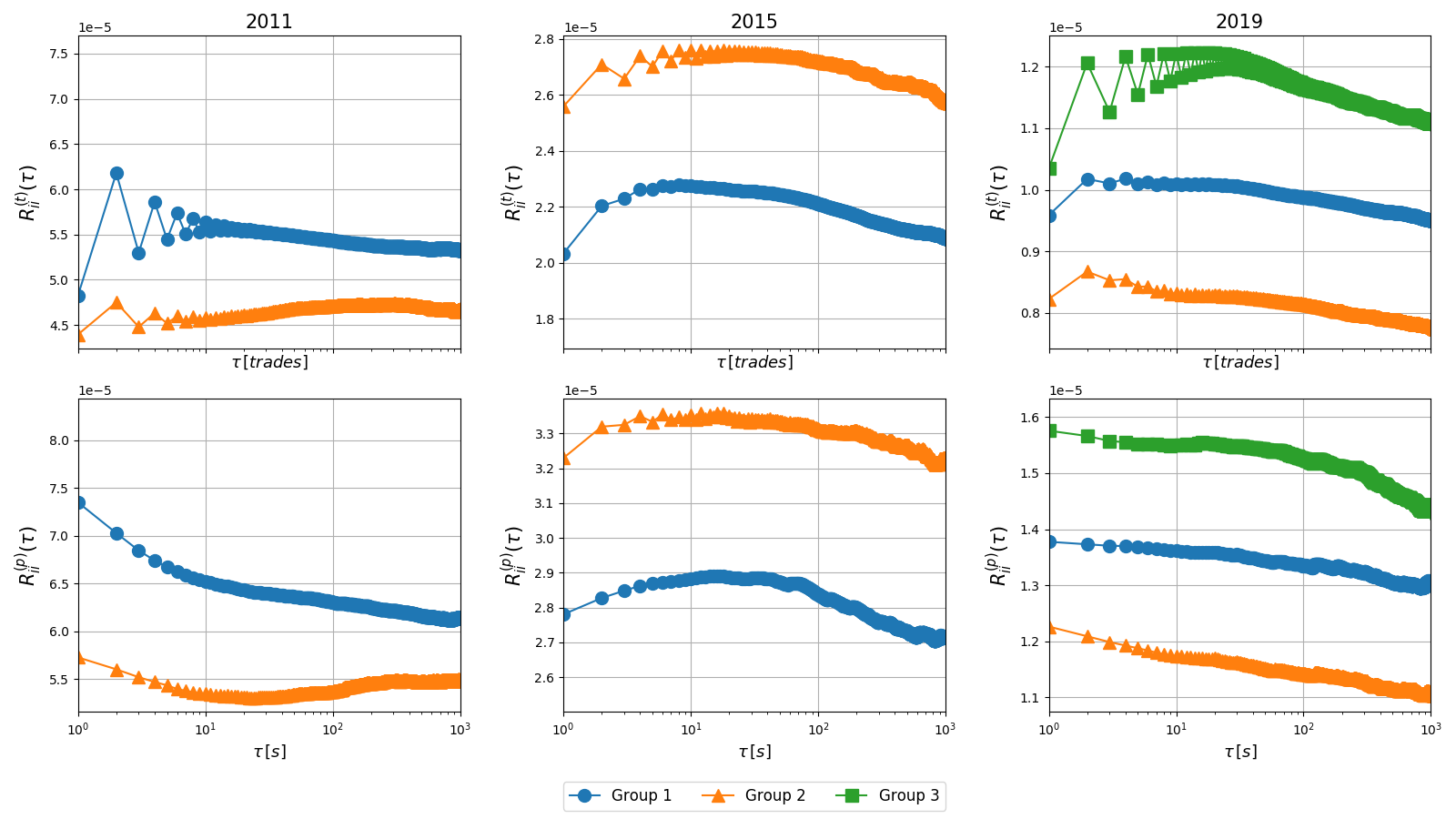}
    \caption{Average price response functions
             $R^{\left(t\right)}_{i}\left(\tau\right)$ versus time lag $\tau$
             on a logarithmic scale in trade time scale (Top) and
             $R^{\left(p\right)}_{ii}\left(\tau\right)$ excluding
             $\varepsilon^{\left(p\right)}_{i}\left(t\right) = 0$ versus time
             lag $\tau$ on a logarithmic scale in physical time scale (Bottom)
             for 47 foreign exchange pairs divided in representative groups in
             three different years (2011, 2015 and 2019).}
    \label{fig:spread_impact}
\end{figure*}

Depending on the year, we identify different numbers of groups according to the pip spread $s_{\textrm{pip}}$. For the years 2011 and 2015, we use two
intervals to select the foreign exchange pairs groups ($s_{\textrm{pip}}<10$
and $10 \le s_{\textrm{pip}}$). In the year 2019 we use three intervals to
select the foreign exchange pairs groups ($s_{\textrm{pip}}<4$,
$4 \le s_{\textrm{pip}} < 10$ and $10 \le s_{\textrm{pip}}$). The detailed
information of the foreign exchange pairs, spread and the groups can be seen in
\ref{app:fx_pairs_spread}. With the groups of the stocks defined, we
average the price response functions of each group.

In Fig. \ref{fig:spread_impact} we show the average response functions for
the corresponding groups in three different years. From year to year the groups
can vary depending on the pip spread. The average price response function for
the pairs with smaller pip spreads (more liquid) have on average the weakest
signal in the figure for all the years and both time scales. On the other hand,
the average price responses for the pairs with larger pip spread (less liquid)
have on average the strongest signal for all the years in both time scales.

From Sect. \ref{sec:response_functions} we expect the increase-maximum-decrease
behavior. This behavior can be seen in the figures of the year 2015 for both
time scales and in the figure of the year 2019 in trade time scale. In these
figures the average price response functions follow an increase, reach a
maximum and then start to slowly decrease. For the other figures, on average,
the response functions start to decrease from the beginning.

The response in trade time scale seems to be noisier, with large changes in the
first time steps. This aggregate noise can be related with the crosses and
exotics pairs, who tend to fluctuate more.

For the years 2015 and 2019 in physical time scale, the
increase-maximum-decrease behavior is not that well defined as in Sect.
\ref{sec:response_functions} or as in the average response in trade time scale.
However, some groups tend to behave in the expected way. The groups that do not
follow the trend, seem to have an instantaneous high response that slowly
decreases with time. This behavior is mostly noticeable in the year 2019.

For the year 2011 in physical time scale, the increase-maximum-decrease shape
is not present. For both groups the response decreased almost immediately.

In the three plots, the foreign exchange market seems to have a global
influence over all the pairs in the corresponding years. A similar behavior
can be seen in correlated financial markets
\cite{my_paper_response_financial}.

%% file: sections/09_conclusion.tex
\section{Conclusion}\label{sec:conclusion}

Price response functions provide quantitative information on the deviation from
Markovian behavior. They measure price changes resulting from execution of
market orders. We used these functions in big data analysis for spot foreign
exchange markets. Such a study was, to the best of our knowledge, never done
before.

We analyzed price response functions in spot foreign exchange markets for
different years and different time scales. We used trade time scale and
physical time scale to compute the price response functions for the seven major
foreign exchange pairs for three different years. These major pairs are highly
relevant in the dynamics of the market. The use of different time scales and
calendar years in the work had the intention to display the different behaviors
the price response function could take when the time parameters differ.

The price response functions were analyzed according to the time scales. On
trade time scale, the signals were noisier. For both time scales we observe
that the signal for all the pairs increases to a maximum and then starts to
slowly decrease. However, for the year 2008 the shape of the signals is not as
well defined as in the other years. The increase-decrease behavior observed in
the spot foreign exchange market was also reported in correlated financial
markets \cite{my_paper_response_financial,Wang_2016_avg}. These results show
that the price response functions conserve their behavior in different years
and in different markets. The shape of the price
response functions is qualitatively explained considering an initial increase caused by the autocorrelated
transaction flow. To assure diffusive prices, price response flattens due to
market liquidity adapting to the flow in the initial increase.

On both scales, the more liquid pairs have a smaller price response function
compared with the non-liquid pairs. As the liquid pairs have more trades during
the market time, the impact of each trade is reduced. Comparing years and
scales, the price response signal is stronger in past than in recent years. As
algorithmic trading has gained great relevance, the quantity of trades has
grown in recent years, and in consequence, the impact in the response has
decreased.

Finally, we checked the pip spread impact in price response functions for three
different years. We used 46 foreign exchange pairs and grouped them depending
on the conditions of the corresponding year analyzed. We employ the year
average pip spread of every pair for each year. For all the year and time
scales, the price response function signals were stronger for the groups of
pairs with larger pip spreads and weaker for the group of pairs with smaller
spreads. For the average of the price response functions, it was only possible
to see the increase-maximum-decrease behavior in the year 2015 in both scales,
and in the year 2019 on trade time scale. Hence, the noise in the cross and
exotic pairs due to the lack of trading compared with the majors seems
stronger. A general average price response behavior for each year and time
scale was spotted for the groups, suggesting a market effect on the foreign
exchange pairs in each year.

Comparing the response functions in stock and spot currency exchange markets
from a more general viewpoint, we find a remarkable similarity. It triggers the
conclusion that the order book mechanism generates in a rather robust fashion
the observed universal features in these two similar, yet different subsystems
within the financial system.

%% file: sections/10_paper_contributions.tex
\section{Author contribution statement}

TG proposed the research. JCHL developed the method of analysis. The idea to
analyze the spread impact was due to JCHL. JCHL carried out the analysis. Both
authors contributed equally to analyzing the results and writing the paper.

    One of us (JCHL) acknowledges financial support from the German Academic
    Exchange Service (DAAD) with the program ``Research Grants - Doctoral
    Programmes in Germany'' (Funding programme 57381412).

    We thank an unknown referee of an earlier version for helpful comments on
    the shape of the response functions.

%% file: sections/11_appendix_A.tex
\section{Foreign exchange pairs used to analyze the spread impact}
\label{app:fx_pairs_spread}

We analyze the spread impact in the price response functions for 47 foreign
exchange pairs in the foreign exchange market for the years 2011, 2015 and
2019. In Table \ref{tab:spread_comp}, we list the pairs in their corresponding
pip spread groups.

\begin{table*}
\begin{center}
\begin{centering}
    \begin{threeparttable}
    \caption{Foreign exchange pairs used in Sect. \ref{sec:spread_impact}.}
    \label{tab:spread_comp}
    \begin{tabular}{lllllll}
    \hline
    Symbol & Pair & Category & Scaling factor & $2011$ & $2015$ & $2019$\tabularnewline
    \hline
    AUD/CAD & Australian dollar/Canadian dollar & Cross & 10000 & G1 & G1 & G1\tabularnewline
    AUD/CHF & Australian dollar/Swiss franc & Cross & 10000 & G1 & G1 & G1\tabularnewline
    AUD/JPY & Australian dollar/Japanese yen & Cross & 100 & G1 & G1 & G1\tabularnewline
    AUD/NZD & Australian dollar/New Zealand dollar & Cross & 10000 & G2 & G1 & G1\tabularnewline
    AUD/USD & Australian dollar/U. S. dollar & Major & 10000 & G1 & G1 & G1\tabularnewline
    CAD/CHF & Canadian dollar/Swiss franc & Cross & 10000 & G1 & G1 & G1\tabularnewline
    CAD/JPY & Canadian dollar/Japanese yen & Cross & 100 & G1 & G1 & G1\tabularnewline
    CHF/JPY & Swiss franc/Japanese yen & Cross & 100 & G1 & G1 & G1\tabularnewline
    EUR/AUD & euro/Australian dollar & Cross & 10000 & G2 & G1 & G1\tabularnewline
    EUR/CAD & euro/Canadian dollar & Cross & 10000 & G1 & G1 & G1\tabularnewline
    EUR/CHF & euro/Swiss franc & Cross & 10000 & G1 & G1 & G1\tabularnewline
    EUR/CZK & euro/Czech koruna & Exotic & 10000 & G2 & G2 & G3\tabularnewline
    EUR/GBP & euro/British pound & Cross & 10000 & G1 & G1 & G1\tabularnewline
    EUR/HUF & euro/Hungarian forint & Exotic & 100 & G2 & G2 & G3\tabularnewline
    EUR/JPY & euro/Japanese yen & Cross & 100 & G1 & G1 & G1\tabularnewline
    EUR/NOK & euro/Norwegian krone & Exotic & 10000 & G2 & G2 & G3\tabularnewline
    EUR/NZD & euro/New Zealand dollar & Cross & 10000 & G2 & G1 & G2\tabularnewline
    EUR/PLN & euro/Polish zloty & Exotic & 10000 & G2 & G2 & G3\tabularnewline
    EUR/SEK & euro/Swedish krona & Exotic & 10000 & G2 & G2 & G3\tabularnewline
    EUR/TRY & euro/Turkish lira & Exotic & 10000 & G2 & G2 & G3\tabularnewline
    EUR/USD & euro/U. S. dollar & Major & 10000 & G1 & G1 & G1\tabularnewline
    GBP/AUD & British pound/Australian dollar & Cross & 10000 & G1 & G1 & G2\tabularnewline
    GBP/CAD & British pound/Canadian dollar & Cross & 10000 & G2 & G1 & G2\tabularnewline
    GBP/CHF & British pound/Swiss franc & Cross & 10000 & G2 & G1 & G1\tabularnewline
    GBP/JPY & British pound/Japanese yen & Cross & 100 & G1 & G1 & G1\tabularnewline
    GBP/NZD & British pound/New Zealand dollar & Cross & 10000 & G2 & G1 & G2\tabularnewline
    GBP/USD & British pound/U. S. dollar & Major & 10000 & G1 & G1 & G1\tabularnewline
    NZD/CAD & New Zealand dollar/Canadian dollar & Cross & 10000 & G1 & G1 & G2\tabularnewline
    NZD/CHF & New Zealand dollar/Swiss franc & Cross & 10000 & G2 & G1 & G1\tabularnewline
    NZD/JPY & New Zealand dollar/Japanese yen & Cross & 100 & G1 & G1 & G1\tabularnewline
    NZD/USD & New Zealand dollar/U. S. dollar & Major & 10000 & G1 & G1 & G1\tabularnewline
    SGD/JPY & Singapore dollar/Japanese yen & Exotic & 100 & G1 & G1 & G1\tabularnewline
    USD/CAD & U. S. dollar/Canadian dollar & Major & 10000 & G1 & G1 & G1\tabularnewline
    USD/CHF & U. S. dollar/Swiss franc & Major & 10000 & G1 & G1 & G1\tabularnewline
    USD/CZK & U. S. dollar/Czech koruna & Exotic & 10000 & G2 & G2 & G3\tabularnewline
    USD/DKK & U. S. dollar/Danish krone & Exotic & 10000 & G1 & G1 & G2\tabularnewline
    USD/HKD & U. S. dollar/Hong Kong dollar & Exotic & 10000 & G1 & G1 & G2\tabularnewline
    USD/HUF & U. S. dollar/Hungarian forint & Exotic & 100 & G2 & G2 & G3\tabularnewline
    USD/JPY & U. S. dollar/Japanese yen & Major & 100 & G1 & G1 & G1\tabularnewline
    USD/MXN & U. S. dollar/Mexican peso & Exotic & 10000 & G2 & G2 & G3\tabularnewline
    USD/NOK & U. S. dollar/Norwegian krone & Exotic & 10000 & G2 & G2 & G3\tabularnewline
    USD/PLN & U. S. dollar/Polish zloty & Exotic & 10000 & G2 & G2 & G3\tabularnewline
    USD/SEK & U. S. dollar/Swedish krona & Exotic & 10000 & G2 & G2 & G3\tabularnewline
    USD/SGD & U. S. dollar/Singapore dollar & Exotic & 10000 & G1 & G1 & G1\tabularnewline
    USD/TRY & U. S. dollar/Turkish lira & Exotic & 10000 & G2 & G1 & G3\tabularnewline
    USD/ZAR & U. S. dollar/South African rand & Exotic & 10000 & G2 & G2 & G3\tabularnewline
    \end{tabular}

    \begin{tablenotes}
    \item[*] G1 = group 1, G2 = group 2 and G3 = group 3.
    \end{tablenotes}
    \end{threeparttable}
\end{centering}
\end{center}
\end{table*}